\documentclass[a4paper,11pt,twoside]{article}
\usepackage{amsmath,amssymb}
\usepackage[ps,dvips,curve,knot,frame]{xy}

\newcommand{\defeq}{:=}

\newcommand{\N}{\mathbb{N}}
\newcommand{\Z}{\mathbb{Z}}

\newcommand{\R}{\mathbb{R}}
\newcommand{\C}{\mathbb{C}}
\newcommand{\im}{\mathrm{i}}
\newcommand{\tens}{\otimes}

\newcommand{\xd}{\mathrm{d}}
\newcommand{\xD}{\mathcal{D}}
\newcommand{\falg}{\mathcal{C}}  

\DeclareMathOperator{\id}{id}

\newcommand{\mR}{\mathsf{R}}

\newcommand{\cR}{\mathcal{R}}

\DeclareMathOperator{\cou}{\epsilon}
\DeclareMathOperator{\cop}{\Delta}
\DeclareMathOperator{\antip}{\mathrm{S}}

\newcommand{\bint}[1]{[#1]}
\newcommand{\phib}{\bar{\phi}}
\newcommand{\U}[1]{\mathrm{U}(#1)}
\newcommand{\SU}[1]{\mathrm{SU}(#1)}
\newcommand{\SO}[1]{\mathrm{SO}(#1)}

\newcommand{\rxy}[1]{{\begin{xy} 0;<1mm,0mm>:<0mm,1mm>::0;0,#1
\end{xy}}}
\newcommand{\sxy}[1]{{\begin{xy} 0;<0.6mm,0mm>:<0mm,0.75mm>::0;0,#1
\end{xy}}}

\begin{document}
\title{\textbf{The Quantum Geometry of\\ Spin and Statistics}}
\author{Robert Oeckl\footnote{email: oeckl@cpt.univ-mrs.fr}\\ \\
Centre de Physique Th\'eorique, CNRS Luminy,
13288 Marseille, France\\
\medskip
and\\
\medskip
Department of Applied Mathematics and Mathematical Physics,\\
University of Cambridge,
Cambridge CB3 0WA, UK}
\date{DAMTP-2000-81\\
27 January 2001}

\maketitle

\vspace{\stretch{1}}

\begin{abstract}
Both, spin and statistics of a quantum system
can be seen to arise from underlying
(quantum) group symmetries.
We show that the spin-statistics theorem is equivalent to a
unification of these symmetries.
Besides covering the Bose-Fermi case we
classify the corresponding possibilities
for anyonic spin and statistics.
We incorporate the underlying extended concept of symmetry
into quantum field theory in a generalised path integral
formulation capable of handling general braid statistics.
For bosons and fermions the 
different path integrals and Feynman rules
naturally emerge without introducing Grassmann
variables.
We also consider the anyonic example of quons and obtain the path
integral counterpart to the usual canonical approach.
\end{abstract}

\vspace{\stretch{1}}
\begin{flushleft}
keywords: quantum groups, statistics, spin, quantum field theory,
 anyons, quons
\end{flushleft}
\section{Introduction}

While spin in quantum physics arises from the geometry of
space-time, statistics is connected to the
geometry of configuration space.
Half-integer spin and
Bose-Fermi statistics arise in 3 or higher dimensions, while in 2
dimensions more general fractional spin and
anyonic statistics are possible
(see \cite{Wil:fracstatany} and references therein). 
In fact, both, spin and statistics are related to symmetries. In the
case of spin this is plainly understood in terms of the symmetries of
space-time.
In the case of statistics the link is more indirect.
From the geometry of a configuration space of identical particles
\cite{LeMy:idparticles} one
is led (in the general case) to the braid group, which acts on it by
particle exchange \cite{Wu:qstatistics}.
From the representation theory of the braid group one naturally arrives
at the concept of braided categories (see, e.g., \cite{Maj:qgroups}) to
describe statistics.
While foreign to ordinary quantum field theory,
such a general formulation of statistics has already been
incorporated into algebraic quantum field theory
\cite{FrReSc:superselbraid, FrGa:braidstatlocal}.
Going further, a reconstruction theorem of quantum group theory tells
us that (essentially) every braided category is the
category of representations of a quantum group. Thus, for any braid
statistics there is a quantum group symmetry of the theory that
generates the statistics. The relevant quantum groups for anyonic
statistics are known \cite{Maj:anyonicqg, Maj:qgroups}.

After reviewing these facts,
we ask, in the first part of the paper (Section~\ref{sec:geospinstat}),
the natural question of whether
and how the (quantum) 
group symmetries behind spin and statistics are related. Such a
connection should be expected in the Bose-Fermi case from the
spin-statistics theorem \cite{Fie:reltheospin, Pau:spinstat}.
In this case, both groups
generating spin and statistics turn out to be (essentially)
$\Z_2$. Remarkably, the statement
of the spin-statistics theorem is found to be precisely equivalent to
the requirement
that the groups be identified. This leads to a quantum symmetry group
that encodes (a) the space-time symmetries, (b) the statistics, and
(c) the spin-statistics theorem. Technically, this quantum group is
the ordinary space-time symmetry (e.g., Poincar\'e) group as a Hopf
algebra, but 
equipped with a non-trivial coquasitriangular structure.
We proceed to explore the possible relations of spin and statistics in
the more general case of fractional spin and anyonic statistics. This
amounts (under certain restrictions) to a classification of all possible
spin-statistics theorems
which could be implemented by a unified quantum group symmetry.

It is essential for our treatment to work
with quantum groups of function algebra type and not of enveloping
algebra type. The global structure of the (quantum) groups, not
visible in the enveloping algebra setting, is crucial
in the unified description of spin and statistics.
In an enveloping algebra setting one would have to provide the global
information ``by hand'', i.e., by adjoining elements.

The second part of the paper (Section~\ref{sec:braidqft}) addresses the
question of how the description of statistics by braided categories
can be incorporated into ordinary quantum field theory.
This is achieved in a path integral formulation by braided quantum
field theory \cite{Oe:bQFT}, replacing and generalising the formalism
of commuting versus anti-commuting variables.
For Bose-Fermi statistics both the bosonic and the fermionic (Berezin)
path integral as well as the different Feynman rules for bosons and
fermions are recovered.

In view of the results of the first part this means that
the sole input of the relevant symmetry quantum group provides
a quantum field theory not only with desired space-time
symmetries, but also with the right statistics and the correct
spin-statistics relation. This appears to be conceptually simpler and
more natural than the somewhat arbitrary introduction of
anti-commuting variables for fermions.

Finally, we explore the potential of the generalised formalism by
considering an example of anyonic statistics. We study the
``quons'' of Greenberg and others \cite{Gre:partviolstat}. We
translate the canonical quon
relations into a braid statistics and
find that braided quantum field theory provides in
this case the path integral counterpart to the canonical
approach.

It has been notoriously difficult to incorporate generalised
statistics into canonical relations. This is exhibited in the quon
case in the necessary absence of any relations between creation or
annihilation operators for different momenta. This gross deviation
from the usual canonical quantisation approach
indicates in our 
opinion that perhaps canonical relations are not the right way to
describe generalised statistics. Instead, the
braided approach advocated here appears more flexible.

In fact, an example of a non-trivial braid statistics in the context
of braided quantum field theory has been described previously
\cite{Oe:nctwist}. This is the statistics
in a quantum field theory twist-equivalent to that on the
noncommutative space arising in string theory.
It is not an anyonic but rather a symmetric and
momentum dependent statistics.

The paper is organised as follows.
Section~\ref{sec:geospinstat} deals with the symmetries underlying
spin and statistics. Subsections~\ref{sec:spin}--\ref{sec:qgstat}
review the necessary foundations in a coherent fashion leading up to
the main
result in Subsection~\ref{sec:spinstat} on the unification of the
symmetries of spin and statistics.
Section~\ref{sec:braidqft} shows how braided quantum field theory
implements general braid statistics into quantum field
theory. Subsections~\ref{sec:braidpath} and \ref{sec:bosepath}
review the braided path integral and its simplification in the
bosonic case. Subsections~\ref{sec:fermipath} and \ref{sec:fermidiag} 
show how the fermionic path integral as well as the fermionic Feynman
rules are recovered from the fermionic
braiding. Subsection~\ref{sec:quons} finally treats the anyonic
example of quons.

In the following,
the term quantum group is taken to mean coquasitriangular Hopf
algebra. We always work over the complex numbers.
For the general theory of quantum groups and braided categories we
refer to Majid's book \cite{Maj:qgroups} and references therein.
\section{The Symmetries behind Spin and Statistics}
\label{sec:geospinstat}

In this section, after reviewing the emergence of spin and
statistics in quantum mechanics, we discuss the formulation of
statistics in terms of braided categories and quantum groups.
This leads us to a unified description of space-time symmetries and
statistics in terms of quantum group symmetries. A spin-statistics
theorem then precisely corresponds to the identification of space-time
and statistical symmetries.
We describe the familiar Bose-Fermi case and
classify the corresponding
possibilities for anyonic spin and statistics.

\subsection{Spin}
\label{sec:spin}

We start by recalling the geometric origin of spin.
In classical mechanics we require that observable quantities form a
representation of the symmetry group of space-time. In quantum
mechanics it is only required that such a representation is
projective, i.e., it is a representation ``up to a phase''
\cite{Wey:groupsqm}. However,
projective representations of a Lie group are in
correspondence to ordinary representations of its universal covering
group \cite{Bar:rayrep}.

Suppose we have some connected orientable (pseudo-) Riemannian
space-time manifold $M$.
We denote its principal bundle of
oriented orthonormal frames by $(E,M,G)$, where $E$ is the total space
and $G$ the structure group, i.e., the orientation preserving isotropy
group.
If $M$ has signature $(n,m)$ then
$G=\SO{n,m}$.
Let $\tilde{G}$ be the universal covering group of $G$.
Denote by
$(\tilde{E},M,\tilde{G})$ the induced lift of $(E,M,G)$
(assuming no global obstructions).\footnote{Strictly speaking, we
should consider coverings of the
global symmetry group. However, if $M$ is a Riemannian homogeneous
space, $E$ can be identified with the global isometry group and
$\tilde E$ with its universal cover (assuming $M$ to be simply
connected).}
Given a representation of $\tilde{G}$ with label $j$, a field with
spin $j$ is described by a section of the corresponding bundle
associated with $(\tilde{E},M,\tilde{G})$.
If $j$ labels a representation of $G$ itself, we say that the spin is
``integer'', otherwise ``fractional''.
Consider the exact sequence
\begin{equation}
 \pi_1(G)\hookrightarrow \tilde{G} \twoheadrightarrow G ,
\label{eq:cover}
\end{equation}
where $\pi_1(G)$ denotes the fundamental group of $G$. A
representation of $\tilde{G}$ is a representation of $G$ if and only
if the induced action of $\pi_1(G)$ is trivial. Thus, loosely
speaking, the ``fractions'' of spin are labeled by the irreducible
representations of $\pi_1(G)$.
In our present context (we assume at most one time direction) there
arise only two different cases which we 
review in the following.

Let $M$ be 3-dimensional Euclidean
space. Then $G=\SO3$ and the exact sequence (\ref{eq:cover}) becomes
\begin{equation}
 \Z_2\hookrightarrow \SU2 \twoheadrightarrow \SO3 .
\label{eq:zsuso}
\end{equation}
With the usual conventions, irreducible representations of $\SU2$ are
labeled by half-integers and those with an integer label descend to
representations of $\SO3$.
$\Z_2$ has just two inequivalent irreducible representations that
distinguish between the two choices, integer or non-integer.
More generally, $\pi_1(\SO{1,n})=\pi_1(\SO{n})=\Z_2$ for all $n\ge 3$.
Thus, if the dimension of space is $\ge 3$ we can only have integer
and half-integer spins.

Now, let $M$ be 2-dimensional Euclidean space. We obtain the exact
sequence
\begin{equation}
 \Z\hookrightarrow \R \twoheadrightarrow \SO2 .
\label{eq:zrso}
\end{equation}
Since the groups are abelian, their unitary irreducible
representations form themselves (abelian) groups. In fact these are
$\SO2$, $\R$, and $\Z$. (The sequence (\ref{eq:zrso})
is Pontrjagin self-dual.) Thus, the unitary irreducible representations
of $\R$ are labeled by real numbers and descend to $\SO2$ if the
label is integer. The ``fractional'' part is labeled by $\U1=\SO2$.
Since also $\pi_1(\SO{1,2})=\Z$, we conclude that in 2 spatial
dimensions continuous real spin is allowed.

Finally, the case of one spatial dimension is degenerate since
the orientation preserving spatial isotropy group is trivial.
We do not discuss this case further.

\subsection{Statistics}
\label{sec:stat}

In the following we review the various possibilities for exchange
statistics arising from the quantisation of a system of identical
particles \cite{LeMy:idparticles, Wu:qstatistics}.
We consider particles in $d$-dimensional Euclidean space. 
Naively, the configuration space for $N$ particles is
$\R^{dN}$. However, due to the particles being identical,
configurations which differ only by 
a permutation of the particle positions are to be considered identical.
Furthermore, we exclude the singularities arising from the subspace
$D\subset \R^{dN}$ where two or more
particle positions coincide. Thus, the true configuration space is
$(\R^{dN}-D)/S_N$, where $S_N$ denotes the symmetric group acting by
exchanging the particle positions. For more than one particle and more
than one dimension it is multiply connected.

\begin{figure}
\begin{center}
\input{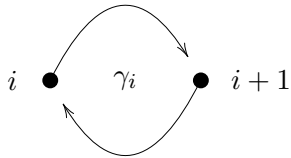}
\caption{Clockwise exchange of particle $i$ with particle $i+1$.}
\label{fig:exchange}
\end{center}
\end{figure}

\begin{figure}
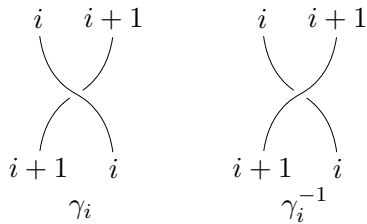

\begin{center}
\begin{tabular}{ccc}
\input{fig_overcross} &&
\input{fig_undercross} \\
$\gamma_i$ && $\gamma_i^{-1}$
\end{tabular}
\caption{Braid generators $\gamma_i$ and $\gamma_i^{-1}$ in
diagrammatic notation.}
\label{fig:cross}
\end{center}
\end{figure}

\begin{figure}
\begin{center}
\input{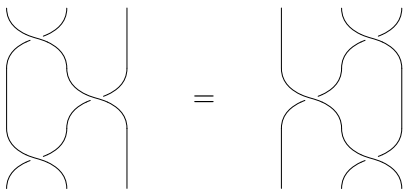}
\caption{Braid relation in diagrammatic notation.}
\label{fig:braidrel}
\end{center}
\end{figure}

Assuming no internal structure for the particles,
quantisation can now be performed by constructing a complex line
bundle with flat connection over
this configuration space. The wave-function is then a
section
of this bundle. If we parallel transport along a non-contractible loop
$\gamma$ in
configuration space, the wave-function picks up a phase factor
$\chi(\gamma)$ coming
from the holonomy of the connection. This defines a one-dimensional
unitary representation of the fundamental group of the configuration
space. (Note that this excludes parastatistics here.)

For dimension $d=2$, the fundamental group of the configuration space
is the braid group on $N$ elements, $B_N$.
It is generated by elements
$\gamma_1,\dots, \gamma_{N-1}$ with relations
$\gamma_i \gamma_j = \gamma_j \gamma_i$ for $i-j\neq\pm 1$ and
\begin{equation}
 \gamma_i\gamma_{i+1}\gamma_i
 =\gamma_{i+1}\gamma_i\gamma_{i+1} .
\label{eq:braidrel}
\end{equation}
To understand this more concretely, consider the
inequivalent ways of exchanging two particles in the plane
without coincidence. They correspond to non-contractible loops
in the configuration space and thus to elements of its fundamental
group. Indeed, $\gamma_i$ corresponds to the exchange of
particle $i$ and particle $i+1$ in (say) clockwise direction, see
Figure~\ref{fig:exchange}. We represent this by a diagram
which can be thought of as depicting the particle trajectories (moving
from top to bottom) as they
exchange, see Figure~\ref{fig:cross}. A counter-clockwise exchange
corresponds to the inverse
$\gamma_i^{-1}$. 
We can also wind the particles around each other more than once,
corresponding to powers of $\gamma_i$ or its inverse.
Figure~\ref{fig:braidrel} shows the
braid relation (\ref{eq:braidrel}) in diagrammatic notation.
The one-dimensional unitary representations of the braid group are
labeled by an angle $\theta$ and take the form
\begin{equation}
 \chi(\gamma_i)=e^{\im\theta}\quad\forall i .
\label{eq:thetastat}
\end{equation}
This was termed $\theta$-statistics in \cite{Wu:qstatistics}.
More generally, a statistics that is induced by representations of the
braid group is called a \emph{braid statistics}.

In dimension $d\ge 3$, the fundamental group of the configuration
space is just the symmetric group $S_N$. It can be obtained from the
braid group by imposing the extra relations
$\gamma_i=\gamma_i^{-1}$. The geometric meaning of this is that the
clockwise and counter-clockwise exchange of particles are
equivalent (homotopic), since we can use the extra dimensions to
deform one path into the other.
Diagrammatically, over- and under-crossings (Figure~\ref{fig:cross})
become
equivalent. The possible representations (\ref{eq:thetastat}) reduce
to just two: bosonic ($\theta=0$) and fermionic ($\theta=\pi$)
statistics.

\subsection{Braided Categories and Statistics}
\label{sec:bcatstat}

\begin{figure}
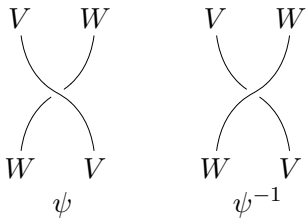

\begin{center}
\begin{tabular}{ccc}
\input{fig_overbraid} &&
\input{fig_underbraid} \\
$\psi$ && $\psi^{-1}$
\end{tabular}
\caption{The braiding and its inverse in
diagrammatic notation.}
\label{fig:braiding}
\end{center}
\end{figure}

\begin{figure}
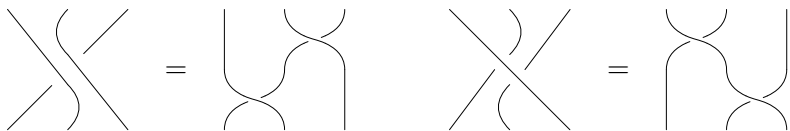

\begin{center}
\begin{tabular}{cccc}
\input{fig_compleft} &&&
\input{fig_compright}
\end{tabular}
\caption{Composition rules for tensor products. Close parallel strands
represent tensor products.}
\label{fig:compose}
\end{center}
\end{figure}

In quantum field theory, multi-particle states are usually expressed in
a Fock space formalism. That is, they are tensor products of one-particle
states. In order to describe a general braid statistics in this
context, we define invertible linear maps
\[
 \psi:V\tens W\to W\tens V
\]
that exchange particles in state spaces $V$ and $W$, and represent the
elements $\gamma_i$ of the braid group. $\psi$ is called a
\emph{braiding}.
We depict
it by the same crossing diagram that we used for $\gamma_i$
(Figure~\ref{fig:cross}). The diagram is now interpreted as
a map, to be read from top to bottom,
where the strands carry the vector spaces $V$ and $W$ respectively
(Figure~\ref{fig:braiding}).
In this formulation, we can easily express the statistics between
different particles as well, by defining $\psi$ for $V$ and $W$ being
different spaces. Also, we are not restricted to one-dimensional
representations of the braid group (or symmetric group) as was the
discussion in Subsection~\ref{sec:stat}. 
Furthermore, we can extend $\psi$ to tensor products
of multi-particle states by composing in the obvious way
(Figure~\ref{fig:compose}).
In fact, we can forget about the origin from the
braid group or symmetric group altogether if we implement the
constraints corresponding to their relations. For the braid group this
is the braid relation, expressed by the diagram in
Figure~\ref{fig:braidrel},
which is now an identity between maps on three-fold tensor
products. For the symmetric group we have the
additional constraint
that $\psi$ and its inverse must be identical, i.e., diagrammatically
over- and under-crossings are identified.

If $\psi$ takes the special form
\begin{equation}
 \psi(v\tens w)=q w\tens v .
\label{eq:qstat}
\end{equation}
with $q\in\C^*$ it is called an \emph{anyonic statistics}, since
particles obeying such statistics are called anyons
\cite{Wil:fracspin}.
If $V$ and
$W$ are state spaces of identical particles without internal structure
we recover $\theta$-statistics
(\ref{eq:thetastat}) with $q=e^{i\theta}$. (Note that we allow
$\theta$ to be complex here.)
The general
expression for Bose-Fermi statistics in this formulation is
\begin{equation}
 \psi(v\tens w)=(-1)^{|v|\cdot |w|} w\tens v ,
\label{eq:bfstat}
\end{equation}
where $|v|=0$ for bosons and $|v|=1$ for fermions.

In fact, what we have described here
is essentially what is called a
\emph{braided category} (see \cite{Mac:categories,Maj:qgroups}).
This means roughly
a collection of
vector spaces closed under the tensor product, together with
a braiding $\psi$ obeying the conditions depicted in
Figure~\ref{fig:compose} and Figure~\ref{fig:braidrel}.
If $\psi=\psi^{-1}$, the braiding and the category are said to be
\emph{symmetric} (since this means restricting to representations of
the symmetric group).
The diagrammatics used here is the standard one
for calculations in braided categories. It arises from
equivalences to categories of
braids, links and tangles \cite{Res:hopfinv,ReTu:ribboninv}.

A description of statistics by braided categories was first employed
in the context of algebraic quantum field theory
\cite{FrReSc:superselbraid, FrGa:braidstatlocal}.
However, it can also be integrated into a (generalised) path integral
formulation of quantum field theory \cite{Oe:bQFT}, as will be
discussed in Section~\ref{sec:braidqft}.

\subsection{Quantum Groups and Statistics}
\label{sec:qgstat}

How do quantum groups come into the game?
We recall a few facts about their representation theory
(see \cite{Maj:qgroups}).
As for ordinary groups, the tensor product $V\tens W$ of
representations $V$, $W$
forms again a representation. However, in contrast to ordinary groups,
the map $\tau:V\tens W\to
W\tens V$ defined by $v\tens w\mapsto w\tens v$ is not in general an
intertwiner, i.e., does not in general commute with the quantum group
action. Instead, for any pair of representations, we are given a
(generally non-trivial) invertible linear
map $\psi:V\tens W\to W\tens V$, which is an intertwiner.
It is encoded in the so called \emph{coquasitriangular structure} of
the quantum group.
In fact, these intertwiners precisely obey the conditions for a
braiding discussed above.
Thus, the category of representations of a
quantum group becomes a braided category.
Remarkably, the converse is also true:
Given a braided category, we can
(under certain technical conditions) reconstruct a quantum group so
that the given category arises as its category of representations.
This is called Tannaka-Krein reconstruction, see \cite{Maj:qgroups}.

Having seen that braided categories can be used to
describe statistics, we can say that
quantum group theory naturally integrates the
notions of ``spin'' (representation theory) and statistics.
More precisely,
the reconstruction theorem tells us that for a given braid
statistics (given by a braided category) there is an underlying
symmetry quantum group, so that the
statistics of particles is determined by their representation
labels.
In the following, we discuss this for anyonic
statistics including the special case of Bose-Fermi statistics.
The relevant quantum groups were identified by Majid
\cite{Maj:anyonicqg, Maj:qgroups} (in a dual formulation of enveloping
algebras).

\begin{table}
\begin{center}
\begin{tabular}{l||c|c|c|c}
quantum group &
$\R$ & $\U1$ & $\Z_n$ & $\Z_2$ \\
\hline
representation labels &
$\R$ & $\Z$ & $\Z_n$ & $\Z_2$ \\
\hline
statistics parameter &
$q\in\C^*$ & $q\in\C^*$ & $q^n=1$ & $q=-1$
\end{tabular}
\caption{Anyonic statistics generating quantum groups.}
\label{tab:statgroups}
\end{center}
\end{table}

The quantum group generating general anyonic statistics turns out to
be the ordinary group $\U1$, but with a non-standard coquasitriangular
structure. As a quantum group it is the function algebra
$\falg(\U1)$. A natural basis in terms of the coproduct are the
Fourier modes $f_k: \phi\mapsto e^{2\pi\im k \phi}$, labeled by
$k\in\Z$. The relations are $f_k f_l=f_{k+l}$,
the coproduct is $\cop f_k=f_k\tens f_k$, the counit is $\cou(f_k)=1$,
and the antipode is $\antip f_k=f_{-k}$. The
coquasitriangular structure $\cR:\falg(\U1)\tens \falg(\U1)\to\C$
that generates the braiding is given by
\begin{equation}
 \cR(f_k\tens f_l)=q^{k l} .
\label{eq:ucqtr}
\end{equation}
The unitary irreducible representations of $\U1$ are labeled by
$k\in\Z$. The braid statistics takes the form
\begin{equation}
 \psi(v_k\tens v_l)=q^{k l} v_l\tens v_k .
\label{eq:ubraid}
\end{equation}
This reduces to expression (\ref{eq:qstat}) for particles in the
representation $k=l=1$.

As it will be of relevance later, we remark that the same anyonic
statistics can also be generated by the group $\R$. As a quantum group
we consider the function algebra $\falg(\R)$ generated by the periodic
functions. The only difference to the $\U1$ case discussed above is
that the basis $\{f_k\}$ is now labeled by real numbers
$k\in\R$ and not just integers. Otherwise the algebraic structure is
the same and
(\ref{eq:ucqtr}) and (\ref{eq:ubraid})
still hold in the same form.
Representations are also labeled by $k\in\R$ now.

For $q=e^{\im\theta}$ an $n$-th root of unity we call the statistics
\emph{rational} since $\theta/2\pi$ is rational. In this case we can
restrict $\U1$ to the subgroup $\Z_n$. This corresponds in the
quantum group setting to the extra relations $f_k=f_{k+n}$, so that we
obtain a finite dimensional quantum group. Irreducible representations
are now labeled by $k\in\Z_n$. However, it will be convenient for the
following discussion to introduce an alternative fractional labeling
by $k\in\frac{1}{n}\Z_n$. Expression (\ref{eq:ubraid}) is thus modified
to
\begin{equation}
 \psi(v_k\tens v_l)=q^{n^2 k l} v_l\tens v_k .
\label{eq:znbraid}
\end{equation}
Expression (\ref{eq:qstat}) is recovered for $k=l=\frac{1}{n}$.
The special case of $\Z_2$ ($\theta=\pi$) is Bose-Fermi statistics
(\ref{eq:bfstat}) with $|v_0|=0$ and $|v_\frac{1}{2}|=1$.

The quantum groups generating anyonic statistics are summarised in
Table~\ref{tab:statgroups} (with the special case of $\Z_2$ listed
separately).

\subsection{Unifying Spin and Statistics}
\label{sec:spinstat}

Having found an underlying ``spin'' connected with
statistics, the natural question arises what possible relation it can
have to the geometric spin discussed in Subsection~\ref{sec:spin}.
This is in essence the question of what spin-statistics theorems are
quantum geometrically realisable. We give a complete answer to this
question in the following (under the restricting assumption that
integer spin particles behave bosonic in dimension $\ge 3$).

We consider first the Bose-Fermi case. 
Our labelling of the $\Z_2$ representations above by $0,\frac{1}{2}$
is already suggestive of an interpretation as the
fractional part of geometric spin.
The latter is also described by
a $\Z_2$, arising from the universal covering of the isotropy group in
dimensions $\ge 3$. And in fact, identifying the two groups
is precisely equivalent to requiring the usual spin-statistics theorem
to hold.

To make this more precise we consider the generic case of
3-dimensional Euclidean space.
Translating the exact sequence (\ref{eq:zsuso}) into quantum group
language, it takes the arrows reversed form\footnote{Note that this is
\emph{not} an exact sequence of vector spaces.}
\begin{equation}
 \falg(\SO3)\hookrightarrow\falg(\SU2)\twoheadrightarrow
 \falg(\Z_2) .
\label{eq:qsosuz}
\end{equation}
Instead of inheriting the trivial coquasitriangular
structure canonically associated to ordinary groups, we equip
$\falg(\Z_2)$ with the non-trivial
coquasitriangular structure generating the Bose-Fermi statistics.
This induces a non-trivial coquasitriangular structure on $\falg(\SU2)$
which precisely exhibits the usual spin-statistics relation.
Explicitly, for a group-like basis $\{1,g\}$ of $\falg(\Z_2)$ and a
Peter-Weyl basis $\{t^l_{ij}\}$, $l\in \frac{1}{2}\N_0$ of
$\falg(\SU2)$, the right hand map of (\ref{eq:qsosuz}) is
$t^l_{ij}\mapsto \delta_{ij} g^{2l}$.
The coquasitriangular structure $\cR(g\tens g)=-1$ on $\falg(\Z_2)$
pulls back to
\[
 \cR(t^l_{ij}\tens t^m_{kl})
 =(-1)^{4lm}\delta_{ij}\delta_{kl}
\]
on $\falg(\SU2)$. This induces the braiding
\[
 \psi(v_k\tens v_l)=(-1)^{4 k l} v_l\tens v_k ,
\]
relating spin and statistics for bosons and fermions in the familiar
way.
The analogous construction can be made in any space-time with spatial
dimension $\ge 3$, since the only relevant input is that the fundamental
group of the isotropy group is $\Z_2$.
This ensures that the function algebra of the covering group is
$\Z_2$-graded into 
functions that are symmetric or antisymmetric with respect to exchange
of the sheets. This decomposition is also a decomposition into
subcoalgebras. Thus, the covering group admits the coquasitriangular
structure
\[
 \cR(f \tens g) =(-1)^{|f|\cdot |g|}\cou(f)\cou(g) .
\]
Note that we can further pull the coquasitriangular
structure back to the relevant global space-time symmetry group by the
same argument. For a treatment of
Bose-Fermi statistics in 2 dimensions see the discussion below.

We now proceed to the more complicated case of
anyonic statistics. Although we can embed $\U1$ into $\SU2$, the
coquasitriangular structure (\ref{eq:ucqtr}) does not pull back from
$\falg(\U1)$ to
$\falg(\SU2)$. The same is true for the other
spin-groups. (This is easily seen by embedding through an intermediate
$\SU2$.) Consequently, we cannot relate the statistical ``spin'' of
anyonic statistics to geometric spin in dimension 3 or higher. Even
in the rational case this is only possible for $q=\pm 1$, which is
the Bose-Fermi case described above.
Thus, we must restrict to 2 dimensions, where the covering of the
spatial isotropy
group is described by the exact sequence (\ref{eq:zrso}). In contrast
to the Bose-Fermi case the statistical group $\U1$ is different from
the group $\Z$ describing the covering. However, we can use the
group $\R$ to generate the statistics instead (see
Table~\ref{tab:statgroups}) and identify it directly
with the universal cover $\R$ of the isotropy group.
We obtain a spin-statistics relation between
anyonic statistics and continuous geometric spin. However, for
$q\neq 1$ we never
have the property that representations which descend to the
isotropy group have bosonic statistics, i.e., trivial braiding with
all other representations.

We can implement this property, however, if we only consider a finite
covering of the isotropy group. This leads to the exact sequence
\[
 \Z_n\hookrightarrow \SO2\twoheadrightarrow \SO2
\]
instead of (\ref{eq:zrso}).
The spins are restricted from continuous values to $n$-th fractions.
We can now establish a spin-statistics relation by
identifying the $\Z_n$ of the covering with the $\Z_n$ of rational
anyonic statistics. The braiding is the one described by
(\ref{eq:znbraid}). Pullback from $\falg(\Z_n)$ to $\falg(\SO2)$
corresponds to extending the representation labels from
$\frac{1}{n}\Z_n$ to $\frac{1}{n}\Z$. Representations of the
covering $\SO2$ that descend to representations of the covered
$\SO2$ are precisely the ones that are bosonic, i.e., have trivial
braiding with all other representations. $n=2$ is the Bose-Fermi
case.

\begin{table}
\begin{center}
\begin{tabular}{l||c|c|c}
spatial dimension &
2 & 2 & $\ge 3$ \\
\hline
statistics quantum group &
$\R$ & $\Z_n$ & $\Z_2$ \\
\hline
statistics parameter &
$q\in\C^*$ & $q^n=1$ & $q=-1$ \\
\hline
integer spin bosonic &
$-$ & $\surd$ & $\surd$
\end{tabular}
\caption{Possible spin-statistics relations.}
\label{tab:spinstat}
\end{center}
\end{table}

Conversely, we may ask the question what possible statistics can be
attached to the geometric spin, i.e., which coquasitriangular structure
is admitted by the relevant (quantum) group. Its turns out that all
the relevant groups are abelian. A coquasitriangular structure on
the function Hopf algebra of an abelian group is a bicharacter on its
Pontrjagin dual, i.e., its group of unitary irreducible representations.
In dimension 3 or higher,
if we require bosonic statistics for
representations descending to the isotropy group, any braiding must be
induced by $\falg(\Z_2)$ in (\ref{eq:qsosuz}). The dual of $\Z_2$ is
$\Z_2$ and there are only two bicharacters on it: The trivial one
(purely bosonic statistics) and the Bose-Fermi one discussed.
In 2 dimensions, the covering group $\R$ of the isotropy group is
self-dual and any bicharacter corresponds to (\ref{eq:ucqtr}) for some
$q\in\C^*$.
We also see the reason now why we were not able to
induce the braiding from $\Z$: Its dual is $\U1$ which has only the
trivial bicharacter. In 2 dimensions with finite covering the relevant
group is $\Z_n$. It is self-dual and the bicharacters correspond to
the different $n$-th roots of unity leading to rational
anyonic statistics.
Thus, our above discussion has already exhausted the possibilities of
attaching statistics to spin.
Table~\ref{tab:spinstat} gives a summary.

Of course, even in the absence of a spin-statistics relation we can
encode the
space-time symmetries as well as the statistics in terms of
just one symmetry (quantum) group. This is then simply the product of the
two relevant (quantum) groups.


\section{Braid Statistics in Quantum Field Theory}
\label{sec:braidqft}

In this section we show how general braid statistics (in the sense of
Section~\ref{sec:geospinstat}) can be incorporated into quantum field
theory in a path integral formulation.
The framework for this is braided quantum field theory
\cite{Oe:bQFT}, a generalisation of quantum field theory
to braided spaces. 
While the motivation for this generalisation in \cite{Oe:bQFT} was
purely to admit quantum group symmetries, 
the natural interpretation of
the braiding in view of Section~\ref{sec:geospinstat} is that of
particle statistics.

Indeed, we show that the correct bosonic and fermionic path integrals
as well as the different Feynman rules for bosons and fermions
emerge with the braiding (\ref{eq:bfstat}) being the only input.
In particular, this does not require
any other a priori distinction between
bosons and fermions such as, e.g., commuting versus
anti-commuting variables.
As an example of anyonic statistics, we study the quons of Greenberg
and others \cite{Gre:partviolstat}, showing that our framework
provides the path integral counterpart to Greenberg's canonical approach.

\subsection{Braided Path Integral}
\label{sec:braidpath}

\begin{figure}
\begin{center}
\input{fig_bint}
\caption{Braided integer.}
\label{fig:bint}
\end{center}
\end{figure}

\begin{figure}
\begin{center}
\input{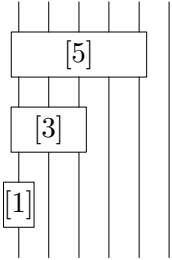}
\caption{The braided double factorial $\bint{5}!!$.}
\label{fig:bdfacex}
\end{center}
\end{figure}

We review the path integral of braided quantum field theory
\cite{Oe:bQFT}.
It is based on the calculus of differentiation and integration in
braided categories developed by Majid \cite{Maj:braidedcalc,
Maj:qgroups} and Kempf
and Majid \cite{KeMa:qintegration}.

Consider Gaussian integration on a finite\footnote{To keep the
discussion less abstract
we restrict it here to the finite dimensional case.} set of variables
$\{\phi_1,\phi_2,\dots\}$. As usual, we define the partition
function\footnote{The Euclidean signature of the action is
chosen for definiteness and does not imply a restriction to Euclidean
field theory.}
\[
Z\defeq\int \xD\phi\, \exp(-S(\phi))
\]
and the (free) $n$-point functions
\begin{equation}
\langle \phi_{k_1} \phi_{k_2}\cdots \phi_{k_n}\rangle
\defeq \frac{1}{Z} \int \xD\phi\, \phi_{k_1} \phi_{k_2}\cdots
\phi_{k_n} \exp(-S(\phi)) ,
\label{eq:corr}
\end{equation}
where $S$ is a quadratic form in $\phi$.
We introduce differentials $\{\partial^1,\partial^2,\dots\}$ dual to
the variables and impose the familiar rules
\begin{gather}
 \int \xD\phi\, \partial^i(\phi_{k_1} \cdots\phi_{k_n} \exp(-S))=0
\label{eq:intdiff}\\
\text{and}\qquad
 \partial^i(\exp(-S))=\partial^i(-S)\exp(-S) .
\label{eq:diffexp}
\end{gather}

For the general case, we assume that the variables satisfy some
definite braid statistics. Recall from Subsection~\ref{sec:bcatstat}
that this means that the vector space $\Phi$ spanned
by $\{\phi_1,\phi_2,\dots\}$ lives in a
braided category. The statistics is encoded in the braiding
$\psi:\Phi\tens\Phi\to\Phi\tens\Phi$, which can be thought of as
representing the exchange of particles. We write more
explicitly
\begin{equation}
 \psi(\phi_i\tens\phi_j)=\sum_{k,l}\mR_{ij}^{kl}\phi_l\tens\phi_k
\label{eq:Rbraiding}
\end{equation}
for a matrix $\mR$. (The braid relation of Figure~\ref{fig:braidrel} is
now equivalent to
the Yang-Baxter equation for $\mR$.)
Under the braiding, the Leibniz rule for
differentiation becomes itself braided. In terms of $\mR$ this can be
expressed by the relation
\begin{equation}
 \partial^i \phi_j = \delta^i_j + \sum_{k,l}\mR_{jk}^{li}\phi_l\partial^k
\label{eq:RLeibniz}
\end{equation}
The $n$-point functions (\ref{eq:corr}) are now completely
determined by the three rules (\ref{eq:intdiff}), (\ref{eq:diffexp}),
and (\ref{eq:RLeibniz}).

It was shown in \cite{Oe:bQFT}
that a braided generalisation of Wick's theorem holds: A $2n$-point function
can be expressed in terms of $n$ $2$-point functions (propagators).
(An $n$-point function for odd $n$ vanishes.)
This is concisely
expressed in the formula
\begin{equation}
 \langle \phi_{k_1} \phi_{k_2}\cdots \phi_{k_{2n}}\rangle
 = \langle \cdot\,\cdot\rangle^n\circ \bint{2n+1}!!
 (\phi_{k_1}, \phi_{k_2},\ldots, \phi_{k_{2n}}) ,
\label{eq:bWick}
\end{equation}
which we explain presently. The symbol $\langle \cdot\,\cdot\rangle^n$
stands for $n$ propagators while $\bint{2n+1}!!$ denotes a
certain linear map from the $2n$-fold tensor product of $\Phi$'s into
itself. Thus,
the right hand side of (\ref{eq:bWick}) means: Take the tensor product
$\phi_{k_1}\tens\cdots\tens\phi_{k_{2n}}$, apply the map
$\bint{2n+1}!!$, then insert the result into $n$ propagators.
The map $\bint{2n+1}!!$ is a composition
\begin{equation}
 \bint{2n-1}!!\defeq (\bint{1}\tens\id^{2n-1})\circ
 (\bint{3}\tens\id^{2n-3}) \circ\cdots\circ (\bint{2n-1}\tens\id),
\label{eq:bdfac}
\end{equation}
called the \emph{braided double factorial}. It is built out of
\emph{braided integers}, which are linear maps defined in
terms of the braiding $\psi$ as
\begin{equation}
 \bint{m}\defeq \id^m+\id^{m-2}\tens\psi^{-1}+\cdots+\psi^{-1}_{1,m-1}
 . \label{eq:bint}
\end{equation}
The diagrammatic interpretation introduced in
Section~\ref{sec:geospinstat}
makes this more transparent. We
represent the braided integer $\bint{m}$ by the linear combination of
diagrams depicted in Figure~\ref{fig:bint}. Each summand of
(\ref{eq:bint}) is
represented by one diagram, containing $m$ strands. As a map, it
is to be read from top
to bottom. Each strand corresponds to one tensor factor of $\Phi$
(i.e., one variable). Crossings
correspond to the braiding $\psi$ or its inverse $\psi^{-1}$
(Figure~\ref{fig:braiding}), while lines that do not cross simply
represent the identity map on that tensor factor.
The composition of maps as in (\ref{eq:bdfac}) is
expressed in terms of diagrams by gluing the strands together, one
diagram on
top of the other. See for example Figure~\ref{fig:bdfacex},
representing the braided double factorial $\bint{5}!!$. Sums of
diagrams are composed by summing over all
compositions of individual diagrams.
One can think of the diagrams as representing the paths of an ensemble
of particles. Each strand is then the track of a particle moving from top
to bottom with crossings corresponding to exchanges.
Alternatively, if one represents the application of the propagators by
connecting the corresponding strands at the bottom one recovers precisely
the usual pictures drawn to illustrate the ordinary Wick theorem.
However, the pictures obtained here
carry additional information encoded in the type of crossing.

We stress that in contrast to ordinary (path) integrals there are no
algebra relations between
the $\phi$'s. The space $\Phi$ just generates a free (noncommutative)
algebra. However,
it is possible to impose relations anyway, provided they are
compatible with the braided differentiation. (More precisely: The
relations must be braided coalgebra maps for the
primitive coproduct on $\Phi$). Such
relations commute with the braided Wick theorem in the sense that
imposing the relations 
first and then evaluating (\ref{eq:bWick}) is the same as evaluating
(\ref{eq:bWick}) first and then imposing the relations.

We restrict now to the case where the braiding
simply permutes
variables with an extra factor. That is, we assume
$\mR_{ij}^{kl}=\alpha_{ij} \delta_i^k\delta_j^l$ in
(\ref{eq:Rbraiding}). Explicitly,
\begin{equation}
 \psi(\phi_i\tens \phi_j)=\alpha_{ij} \phi_j\tens\phi_i .
\label{eq:xstat}
\end{equation}
This is
sufficient for considering bosonic, fermionic, and anyonic statistics.
The braided integers become sums of
permutations equipped with extra factors. Consequently, we can express
the braided Wick theorem (\ref{eq:bWick}) in a more familiar way:
\begin{equation}
\langle \phi_{k_1} \phi_{k_2}\cdots \phi_{k_{2n}}\rangle
= \sum_{\text{pairings}} \alpha(P) \langle \phi_{k_{P_1}}
\phi_{k_{P_2}}\rangle
\cdots \langle \phi_{k_{P_{2n-1}}} \phi_{k_{P_{2n}}}\rangle ,
\label{eq:sWick}
\end{equation}
where the sum runs over all permutations $P$ of $\{1,\dots,2n\}$
leading to inequivalent pairings. Generically, $\alpha$ is some
complicated function of $P$, built out of the $\alpha_{ij}$. It does
not, in general, define a representation of the symmetric group.
Note also that the order of the variables
in each propagator on the right hand side is relevant. It is
such that the two variables
are in the same order on both sides of the equation.

\subsection{Bosonic Path Integral}
\label{sec:bosepath}

Setting $\alpha_{ij}=1$ in (\ref{eq:xstat}) defines
bosonic statistics. In this case, the braided path integral and the
Feynman rules reduce by construction to the ordinary ones of
bosonic quantum field theory \cite{Oe:bQFT}. 
Nevertheless, we include the bosonic integral here for
completeness and to prepare the ground for the fermionic case.

The Leibniz rule (\ref{eq:RLeibniz}) becomes the ordinary one
$[\partial^i,\phi_j]=\delta^i_j$ and we recover the relevant bosonic
differentiation and integration rules.
In expression (\ref{eq:sWick}) we get $\alpha(P)=1$ and
arrive at the bosonic Wick theorem, which merely expresses the
combinatorics of grouping variables into pairs.
Writing
$
 S(\phi)=\frac{1}{2}\sum_{i,j} \phi_i A_{ij} \phi_j
$
for a symmetric matrix $A$, the resulting propagator
is $\langle \phi_{k} \phi_{l} \rangle = A^{-1}_{k l}$.

As a combinatorial exercise we can count the
number of terms in (\ref{eq:sWick}) by giving each propagator the
numerical value 1. This amounts to replacing the braided
integers in (\ref{eq:bWick}) by ordinary integers. The braided double
factorial turns into an ordinary double factorial and we obtain the
value $(2n-1)!!=(2n)!/(n! 2^n)$, which is
precisely the number of ways in which we can arrange $2n$
variables into pairs of two.

For the case of
conjugated variables $\{\phi_1,\phi_2,\dots\}$ and
$\{\phib_1,\phib_2,\ldots\}$ we require $S$ to have the form
$
 S(\phi)=\sum_{i,j} \phib_i B_{ij} \phi_j
$
for a matrix $B$. This is the same however, as taking all variables
$\{\phi_1,\phi_2,\dots,\phib_1,\phib_2,\dots\}$ together and requiring
$A$ to have the form
\[
 A=\begin{pmatrix} 0 & B^T \\ B & 0 \end{pmatrix} .
\]
The propagator becomes $\langle \phi_{k} \phib_{l} \rangle
= \langle \phib_{l} \phi_{k} \rangle
= B^{-1}_{k l}$ with propagators of two un-barred or two barred
variables vanishing. Consequently, Wick's theorem specialises to its
familiar form for conjugated variables
\begin{equation}
\langle \phi_{k_1} \phib_{l_1}\cdots \phi_{k_n}\phib_{l_n}\rangle
= \sum_{\text{permutations }P} \langle \phi_{k_1}
\phib_{l_{P_1}}\rangle
\cdots \langle \phi_{k_n} \phib_{l_{P_n}}\rangle ,
\label{eq:cWick}
\end{equation}
where the sum runs over all permutations $P$ of
$\{1,\dots,n\}$.

\subsection{Fermionic Path Integral}
\label{sec:fermipath}

Setting $\alpha_{ij}=-1$ in (\ref{eq:xstat}) defines
fermionic statistics. We show that the
resulting path integral is equivalent to the Berezin path integral for
fermions in standard quantum field theory.

The Leibniz rule (\ref{eq:RLeibniz}) becomes $\{\partial^i,\phi_j\}
=\delta^i_j$. This is indeed the familiar expression for Grassmann
variables, which are usually employed to perform the fermionic
integration. Furthermore, the other rules (\ref{eq:intdiff}) and
(\ref{eq:diffexp}) that we have required to define $n$-point
functions turn out to hold
also for Grassmann variables. This is quite obvious for
(\ref{eq:intdiff}), since differentiation and integration are the same
for Grassmann variables and differentiating twice by the same variable
must result in zero. 
Writing
$
 S(\phi)=\frac{1}{2}\sum_{i,j} \phi_i A_{ij} \phi_j
$
with anti-symmetric matrix $A$,
(\ref{eq:diffexp}) follows from the observation that the relations
$[\partial^i,S]=A_{ij}\phi_j$ and $[\phi_i,S]=0$ have the same
commutator form as
in the bosonic case, since $S$ is quadratic. Thus, fermionic
braided integration and integration with Grassmann variables must
agree. 
Indeed, in expression (\ref{eq:sWick})
$\alpha(P)$ becomes the signature of the permutation $P$. This is
Wick's theorem for fermions. Also, the propagator becomes the correct
one $\langle
\phi_{k} \phi_{l} 
\rangle = A^{-1}_{k l}$.

As in the bosonic case, we can play the game to assign each propagator
the numerical value 1. This time we
count the difference between the number of terms contributing
with a plus sign and those with a minus sign in (\ref{eq:sWick}).
In the diagrammatic language, this amounts to replacing any
diagram by $1$ or $-1$ depending on whether it contains an even or odd
number of crossings. For the braided integers this means that
$\bint{m}$ takes the value $1$ if $m$ is odd and zero if $m$ is
even. Since the braided factorial is a product of odd integers it
takes the value $1$, which is the desired result.

For conjugated variables $\{\phi_1,\phi_2,\dots\}$ and
$\{\phib_1,\phib_2,\ldots\}$ we write
$
 S(\phi)=\sum_{i,j} \phib_i B_{ij} \phi_j
$
for a matrix $B$. This is the same as requiring $A$ to have the form
\begin{equation}
 A=\begin{pmatrix} 0 & -B^T \\ B & 0 \end{pmatrix} .
\label{eq:fermconj}
\end{equation}
The propagator becomes $\langle \phi_{k} \phib_{l} \rangle
= -\langle \phib_{l} \phi_{k} \rangle
= B^{-1}_{k l}$ with propagators of two un-barred or two barred
variables vanishing. Similarly to the bosonic case, Wick's theorem
specialises to the form (\ref{eq:cWick}), although with a factor
inserted that takes the value $1$ or $-1$ depending on the signature
of the permutation $P$.

\subsection{Fermionic Feynman Diagrams}
\label{sec:fermidiag}

In braided quantum field theory, no a priori distinction is made in
the treatment of fields with different statistics in (generalised)
Feynman diagrams.
While for bosonic fields the correct Feynman rules are obtained by
construction, this is less apparent for fermionic
fields. 
We show in the following that the correct fermionic Feynman
rules indeed emerge.

\begin{figure}
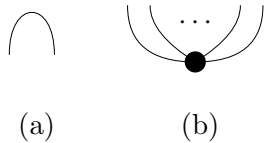

\begin{center}
\begin{tabular}{ccc}
\input{fig_arch} &&
\input{fig_vertex} \\ \\
(a) && (b)
\end{tabular}
\caption{Propagator (a) and vertex (b).}
\label{fig:archvert}
\end{center}
\end{figure}

We briefly recall the rules for composing braided Feynman diagrams,
i.e., the generalised Feynman diagrams of braided quantum field
theory.
This extends the diagrammatic language already introduced.
As usual, diagrams are composed of propagators and vertices. However,
they have to be arranged in a certain way. Propagators, which are
represented by arches (Figure~\ref{fig:archvert}.a), are all drawn
next to each other
at the top of a diagram. Vertices, represented by dots that
collect several strands together (Figure~\ref{fig:archvert}.b) are
drawn next to each
other at the bottom. The propagators are connected with the vertices
by lines which can cross. These crossings are precisely the braiding
or its inverse (Figure~\ref{fig:braiding}).
External lines simply end on the bottom line of the diagram without
meeting a vertex.

For bosons, all crossings are trivial and we recover the
usual bosonic Feynman rules. For fermions, over- and under-crossings
are still identical, but introduce a factor of $-1$.
This is the only difference between bosons and fermions in braided
Feynman diagrams.
At first sight this appears to be at odds with
standard quantum field theory, which prescribes no factor
for line crossings, but introduces extra rules for fermions
instead: (a) Each exchange of external fermion lines
introduces a factor of $-1$, (b) Each internal fermion loop
contributes a factor of $-1$.
In fact, both prescriptions are equivalent as we proceed to show in
the following.

It is easy to see how rule (a) comes about. Exchanging external
fermion lines is achieved by introducing (or removing) crossings. The
number of crossings is necessarily odd, since the exchanged lines
cross once, while any other lines are crossed twice (once by each of
the two which are to be exchanged). Furthermore, crossings of external
lines with loops or of loops with loops do not contribute since they
always appear in pairs. It remains to be shown how rule (b) arises.

\begin{figure}
\begin{center}
\input{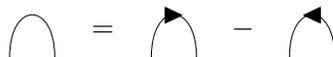}
\caption{Decomposition of the fermionic propagator.}
\label{fig:propconj}
\end{center}
\end{figure}

\begin{figure}
\begin{center}
\input{fig_vertconj}
\caption{Decomposition of fermionic vertices.
The dotted lines (drawn downwards for ease of notation)
represent other fields.}
\label{fig:vertconj}
\end{center}
\end{figure}

\begin{figure}
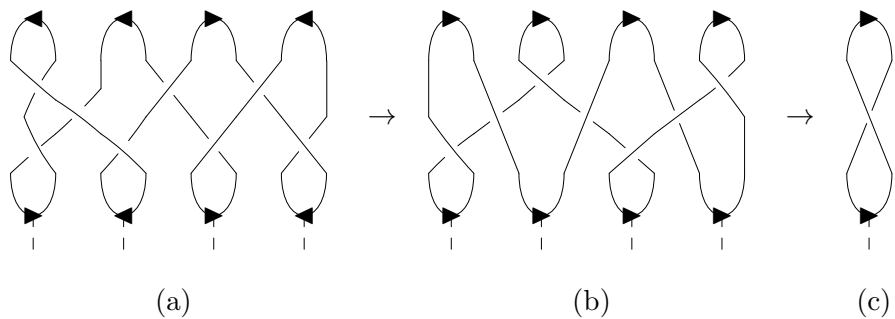

\begin{center}
\begin{tabular}{ccccc}
\input{fig_floop1} & $\rightarrow$ &
\input{fig_floop2} & $\rightarrow$ &
\input{fig_floop3} \\ \\
(a) & & (b) & & (c)
\end{tabular}
\caption{Evaluating the sign of a fermion loop.}
\label{fig:floops}
\end{center}
\end{figure}

First, we note that ordinary fermions are described by conjugated
variables. Thus, the propagator consists of two components
corresponding to the two blocks in (\ref{eq:fermconj}).
Usually, one
picks out one component and indicates which one it is by an arrow
(Figure~\ref{fig:propconj}).
The two components have a relative minus sign as in 
(\ref{eq:fermconj}) due to fermionic anti-symmetry. 
The same applies to fermion vertices
(Figure~\ref{fig:vertconj}).\footnote{Note that the relative choice of
positive 
orientation between propagator and vertex is the choice of sign
for the vertex term in the action.}
Since only matching orientations contribute, each Fermion line
decomposes into two components with consistently chosen orientation
of propagators and vertices.

Consider now a fermion loop (see the example in
Figure~\ref{fig:floops}.a for
illustration). We have to sum over both orientations of the loop in
general. We consider the contribution of one of the two
orientations. Its sign is determined from the various
crossings and orientation choices of the propagators and vertices. To
simplify, we choose the positive orientation and twist any propagators
and vertices with the negative orientation around
(Figure~\ref{fig:floops}.b). This does not alter the sign, since
crossings and
orientation changes are introduced in equal number. Now, the sign
contribution of the diagram is determined by the number of
crossings modulo 2. To find it, we remove propagators and vertices
pair-wise by straightening out lines, keeping in mind that we are
always allowed to change over- 
in under-crossings and vice versa. This removes crossings only
pair-wise and leaves the sign invariant. We are left with just one
propagator and vertex (Figure~\ref{fig:floops}.c). This diagram must have 
one (or an odd number of) crossing. Thus, the overall factor is $-1$,
in agreement with standard quantum field theory.

As a further remark, 
the strict rules for the
arrangement of propagators and vertices in braided Feynman
diagrams can be relaxed to the ordinary rules if the braiding is
symmetric ($\psi=\psi^{-1}$), as is the case for fermions. This is
discussed in \cite{Oe:nctwist}.

\subsection{Anyonic Statistics and Quons}
\label{sec:quons}

In this section, we investigate anyonic statistics, i.e., we are
interested in the case $\alpha_{ij}=q$ for $q\in\C^*$ in
(\ref{eq:xstat}).
Since there is no standard quantum field theory of anyons to compare
with, we start from a canonical approach. This also sheds new
light on the bosonic and fermionic case from this point of view.
More specifically, we consider the ``quons'' which provide an
interesting example
of anyons studied by Greenberg and others \cite{Gre:partviolstat}.

Consider the relations
\begin{equation}
 a_k a_l^\dagger = \delta_{kl} + q a_l^\dagger a_k
\label{eq:quonrel}
\end{equation}
between creation and annihilation operators.
Greenberg's treatment of this algebra is motivated by the
possibility of 
small violations of bosonic ($q=1$) or fermionic ($q=-1$)
statistics. However, we need not take this point of view here.

In contrast to the ordinary canonical approach, no relations among
$a$'s or $a^\dagger$'s are introduced. In fact, such relations
are not needed for normal ordering or the calculation of vacuum expectation
values, as was stressed in \cite{Gre:partviolstat}. We are going one
step further by remarking that relation (\ref{eq:quonrel}) is only ever
evaluated in one direction: from left to right.
Thus, one could interpret (\ref{eq:quonrel}) as defining the
exchange statistics between a particle ($a_l^\dagger$) and a ``hole''
($a_k$), where the $\delta$-term just comes from the operator
picture, analogous to expression (\ref{eq:RLeibniz}). The
corresponding braiding
is
\begin{equation}
 \psi^{-1}(a_k\tens a_l^\dagger)= q a_l^\dagger \tens a_k .
\label{eq:quonbraidmix}
\end{equation}
The choice of $\psi^{-1}$ over $\psi$ is to conform with the
conventions of braided quantum field theory, where only $\psi^{-1}$
appears in (\ref{eq:bdfac}).
In fact, we wish to make the whole Hilbert space of states into a
braided space, in the spirit of Section~\ref{sec:geospinstat}.
In order for expressions with an equal number of creation and
annihilation operators (``zero particle number'') to behave bosonic,
we need to impose
\begin{equation}
 \psi^{-1}(a_k\tens a_l)= q^{-1} a_l \tens a_k
 \quad\text{and}\quad
 \psi^{-1}(a_k^\dagger\tens a_l^\dagger)= q^{-1} a_l^\dagger \tens
 a_k^\dagger .
\label{eq:quonbraidsame}
\end{equation}
The particles and holes obey anyonic statistics among
themselves.

We take the statistics generating group according to
Table~\ref{tab:statgroups} to be
$\U1$. Thus, we have the general expression
\begin{equation}
 \psi(v\tens w)=q^{|v|\cdot |w|} w\tens v
\label{eq:quonbraid}
\end{equation}
for the statistics. Particles are in the representation
$|a_k^\dagger|=1$ and holes in the representation $|a_k|=-1$ so that
we recover (\ref{eq:quonbraidmix}) and (\ref{eq:quonbraidsame}).

A massive real scalar field is expressed in terms of creation and
annihilation operators as
\[
 \phi(x)=\int \frac{\xd^3 k}{\sqrt{(2\pi)^3 2\omega_k}}
  \left(a_k e^{-\im k\cdot x} + a_k^\dagger e^{\im k\cdot x}\right)
\]
with $\omega_k =\sqrt{k^2+m^2}$. We split it as usual into the
components $\phi(x)=\phi^+(x)+\phi^-(x)$, where $\phi^+(x)$ only
contains annihilation operators while $\phi^-(x)$ only contains
creation operators. We can view this formally as a
decomposition of the space of classical fields
$\Phi=\Phi^+\oplus\Phi^-$. The statistics inherited from the canonical
picture is given by the $\U1$ representation labels $|\phi^+(x)|=-1$
and $|\phi^-(x)|=1$. As a remark, we observe that upon reducing $\U1$ to
$\Z_2$ we have $1\cong -1$ as representations. This can be seen to be
the reason why no analogous splitting of the field was necessary in the
fermionic case.

With the braiding defined on the classical field,
the path integral description of the quon is now precisely given
by the path integral of braided quantum field theory. 
And indeed, the braided Wick theorem (\ref{eq:bWick}) specialises in this
case to the one found by Greenberg in \cite{Gre:partviolstat}.
We consider the example of the free 4-point function. Its
decomposition into propagators is given by
\begin{equation}
\begin{split}
 & \langle \phi(x_1)\phi(x_2)\phi(x_3)\phi(x_4)\rangle
 =\langle \phi(x_1)\phi(x_2)\rangle\langle\phi(x_3)\phi(x_4)\rangle \\
 & +q \langle \phi(x_1)\phi(x_3)\rangle\langle\phi(x_2)\phi(x_4)\rangle
 + \langle \phi(x_2)\phi(x_3)\rangle\langle\phi(x_1)\phi(x_4)\rangle .
\label{eq:quon4pt}
\end{split}
\end{equation}
This reproduces (37--39) in \cite{Gre:partviolstat}.\footnote{
Greenberg uses a complex scalar field.
However, it is clear how to obtain (37--39) in \cite{Gre:partviolstat}
from (\ref{eq:quon4pt}):
Just insert the $\dagger$'s and remove propagators that are not pairs
of a $\phi$ and a $\phi^\dagger$.}

\begin{figure}
\begin{center}
\input{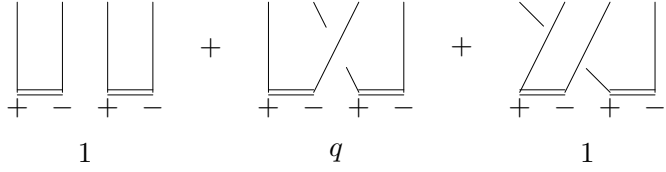}
\caption{The contributions to the quon 4-point function.}
\label{fig:fourpoint}
\end{center}
\end{figure}

To see how (\ref{eq:quon4pt}) comes about consider
Figure~\ref{fig:fourpoint}.
The braided double factorial $\bint{3}!!=\bint{3}\tens\id$ produces
a sum of three diagrams. At the bottom we have indicated by horizontal
double lines the evaluation by the propagators. In order to see what
factors the braidings introduce we note that only the combination
$\langle \phi^+(x)\phi^-(y)\rangle$ makes a contribution to the
propagator. Accordingly, we have written below each strand the sign
indicating the relevant field component carried by the strand.
The evaluation is now simply determined by the statistics of the
relevant field components: A braiding of a $+$ with a $-$
field gives a factor of $q$ while braidings among $+$ or $-$ fields
give a factor of $q^{-1}$.
In this way, any free $n$-point function can be easily evaluated. Note
that the rule for obtaining the $q$-factors given by Greenberg appears
to be slightly different,
but is equivalent. If, while fixing the attachments of the strands at
the top line we deform the strands (with the attached propagators) so
as to minimise the number of intersections, we are only left with
intersections between fields with different sign labels. These all
introduce factors of $q$. This is Greenberg's description.

Finally, we consider the issue of the statistics of bound states of
quons. It was found in \cite{GrHi:qstatcomp} that bound states of $n$
quons have a statistics parameter of $q^{n^2}$. In fact, in our
formulation this follows from the knowledge of the
(quantum) group symmetry behind the statistics. A quon and its
creation operator is in a $1$-representation of the statistics
generating $\U1$. A quon hole and the annihilation operator
are in the $-1$-representation. Thus, an $n$-quon state
or operator that increases the quon number by $n$ lives in an
$n$-representation. By formula (\ref{eq:quonbraid}) we find
that the statistics factor between two such objects is $q^{n\cdot n}$.

Although usually considered in the
context of small violations of Bose or Fermi statistics in higher
dimensions, our analysis suggests that it would be quite natural to
consider quons in 2 dimensions where a spin-statistics relation can be
established quantum-geometrically
as shown in Subsection~\ref{sec:spinstat}.

\section*{Acknowledgements}

I would like to thank Shahn Majid, Hendryk Pfeiffer and Fabian Wagner
for stimulating discussions.
I would also like to acknowledge financial support by
the German Academic Exchange Service (DAAD) and
the Engineering and Physical Sciences Research Council (EPSRC).

\appendix
\bibliographystyle{amsordx}
\bibliography{stdrefs}
\end{document}